\documentclass[10pt]{iopart}

\usepackage{iopams}  
\usepackage{epstopdf}
\usepackage{graphicx}
\begin{document}

\title[Coherent gigahertz phonons in Ge$_{2}$Sb$_{2}$Te$_{5}$ phase-change materials]{Coherent gigahertz phonons in Ge$_{2}$Sb$_{2}$Te$_{5}$ phase-change materials}

\author{Muneaki Hase$^1$, Paul Fons$^2$, Alexander V. Kolobov$^2$, Junji Tominaga$^2$}
\address{$^1$Division of Applied Physics, Faculty of Pure and Applied Sciences, University of Tsukuba, 1-1-1 Tennodai, Tsukuba 305-8573, Japan}
\address{$^2$Nanoelectronics Research Institute, National Institute of Advanced Industrial Science and Technology (AIST), 
Tsukuba Central 4, 1-1-1 Higashi, Tsukuba 305-8562, Japan}
\ead{mhase@bk.tsukuba.ac.jp}

\vspace{10pt}
\begin{indented}
\item[]Received September, 2015
\end{indented}

\begin{abstract}
Using $\approx$40 fs ultrashort laser pulses, we investigate the picosecond acoustic response from a prototypical phase change material, thin Ge$_{2}$Sb$_{2}$Te$_{5}$ (GST) films with various thicknesses. After excitation with a 1.53 eV-energy pulse with a fluence of $\approx$ 5 mJ/cm$^{2}$, the time-resolved reflectivity change exhibits transient electronic response, followed by a combination of exponential-like strain and coherent acoustic phonons in the gigahertz (GHz) frequency range. The time-domain shape of the coherent acoustic pulse is well reproduced by the use of the strain model by Thomsen {\it et al} (Phys. Rev. B 34, 4129, 1986). We found that the decay rate (the inverse of the relaxation time) of the acoustic phonon both in the amorphous and in the crystalline phases decreases as the film thickness increases. The thickness dependence of the acoustic phonon decay is well modeled based on both phonon-defect scattering and acoustic phonon attenuation at the GST/Si interface, and it is revealed that those scattering and attenuation are larger in crystalline GST films than those in amorphous GST films. 
\end{abstract}

%\pacs{78.47.J-, 43.35.+d, 65.60.+a, 63.20.kp}
\maketitle

\section{Introduction}
Phase change data storage technology offers high speed, rewritable, 
and reliable nonvolatile solid state memory, which may overcome the problems of the current generation of 
Si-based memory technologies \cite{Ovshinsky68}. In phase change memory (PCM) materials, the switching 
between a high resistance amorphous and low resistance crystalline phases can be 
carried out by optical means. 
One of the most common and reliable materials for modern optical recording is 
Ge$_{2}$Sb$_{2}$Te$_{5}$ (GST), in which the phase transition between the crystalline and 
amorphous phases governs rewritable recording \cite{Yamada91}. Recently, extensive 
theoretical investigations regarding the mechanism of the phase change in GST have been carried out using 
molecular dynamics simulations \cite{Akola07,Hegedus08,Li11}. In addition, experimental studies 
using extended 
x-ray absorption fine structure (EXAFS) \cite{Kolobov04}, time-resolved x-ray absorption near-edge 
structure (XANES) \cite{Fons10} and Raman scattering measurements \cite{Andrikopoulos07,Krbal11} have 
examined the local atomic arrangements in GST materials. 
Knowledge of thermal properties is important to optimize the performance of phase change devices \cite{Lyeo06}, 
whereas lower thermal conductivity enables one to realize low power switching operation, where intense 
laser irradiation causes lattice heating \cite{Lee09,Wong10}. 
Among the lattice thermal properties, however, the athermal component of amorphization in phase change materials is 
a particularly important issue \cite{Kolobov11} because it enables us to maintain the switching speed and 
energy required for the phase change. 

Femtosecond optical pump-probe spectroscopy is a powerful tool to investigate the athermal component of either 
amorphization or crystallization in phase change materials and it has been extensively used for understanding  
sub-nanosecond fast phase change dynamics \cite{Siegel04,Siegel08,Siegel12}.
Coherent phonon spectroscopy (CPS) \cite{Cheng90,Dekorsy00} has recently been applied to 
GST materials of alloyed polycrystalline \cite{Forst00}, epitaxial \cite{Shalini15} and superlattice films \cite{Hase09,Makino11,Hase15}, and related Sb$_{2}$Te$_{3}$ 
films \cite{Li10}. In these studies, the 
observed optical phonon modes in the amorphous GST films were found to be strongly damped 
modes, with a relaxation time of less than a few picoseconds due to scattering by lattice 
defects \cite{Forst00,Hase09}. CPS on GST compounds, however, has rarely been applied to investigate 
thermal properties by means of the generation of acoustic phonons and has only been applied for a limited number 
of conditions and focuses \cite{Kim13,Shu13}. 

Since the next generation phase change devices may be produced as thin deposited film on silicon substrates \cite{Raoux08,Simpson10}, it is important to study the thermal properties, such as the sound velocity and the attenuation of acoustic phonons in phase change materials, whose thickness is less than several tens of nanometers. 
Moreover, the focus of recent research on the nature of disorder-induced localization in crystalline phase has been on a metal-insulator transitions, where Anderson localization has been considered to play a major role \cite{Siegrist11}. Therefore, investigations of disorder, e.g., grain boundaries, lattice defects (vacancies, dislocations and impurities), are highly important to understand the nature of localization in phase change materials. 
 
To investigate the effect of the film thickness on the degree of disorder, we present a time-domain study of picosecond acoustic lattice dynamics in a prototypical phase change material, a Ge$_{2}$Sb$_{2}$Te$_{5}$ film. After excitation with 1.53 eV-energy and a fluence of $>$ 1 mJ/cm$^{2}$ photons, the time-resolved reflectivity exhibits 
a transient electronic response, followed by a combination of exponential-like strain and a traveling acoustic strain pulse in the GHz frequency range. The acoustic stress model well describes the experimental data, while electronic stress dominates the generation of acoustic phonon in the amorphous films. 
Moreover, the difference in the time scale of the acoustic phonon relaxation found in GST films as a function of the film thickness is used to ascertain the physical origin of acoustic phonon scattering in Ge$_{2}$Sb$_{2}$Te$_{5}$ and at the GST/Si interface, and we demonstrate that phonon scattering dynamics are different in amorphous and crystalline GST films. 

\section{Methods}
\subsection{Experimental technique}
The samples used in the present study were thin films of Ge$_{2}$Sb$_{2}$Te$_{5}$ fabricated 
using a helicon-wave RF magnetron sputtering on a Si (100) substrate. The thickness $d$ of the films was varied for $d$ = 20, 30, 40, and 80 nm in order to investigate the relation between the absorption length and acoustic phonon propagation. 
By annealing the amorphous Ge$_{2}$Sb$_{2}$Te$_{5}$ (amo-GST) films at 180 $^{\circ}$C for more than 1 hour, the crystalline (face-center-cubic: fcc) Ge$_{2}$Sb$_{2}$Te$_{5}$ phase (fcc-GST) was also obtained \cite{Forst00,Tominaga08}. 

A reflection-type pump-probe measurement was carried out at room temperature. Femtosecond seed pulses from a Ti:sapphire laser oscillator, 
operating at a wavelength of 810 nm with a pulse duration of about $\approx$20 fs, were amplified to a pulse energy of 6 $\mu$J in a 100 kHz regenerative-amplifier 
system. After compensation for the amplifier dispersion, $\approx$40 fs duration pulses with a central wavelength of $\lambda$ = 810 nm ($E$ $\approx$ 1.53 eV) were obtained. 
The pump and probe beams were focused on the samples to a diameter of $\sim$ 100  and $\sim$ 50 $\mu$m, respectively. 
The pump fluence was varied with a neutral density filter between 1.3 mJ/cm$^{2}$ and 5 mJ/cm$^{2}$, and the sample was not exposed to fluences higher than 16 mJ/cm$^{2}$ to prevent photo-bleaching effects and/or to cause permanent phase change \cite{Rueda11,Takeda14,Fons14}, and throughout the experiment these conditions were maintained. 
The probe pulse energy was also reduced and fixed at 0.1 mJ/cm$^{2}$. 
The transient reflectivity (TR) change $\Delta R/R$ was recorded by scanning the time delay over a time range up to 120 ps and averaged for 1000 scans by using a shaker operated a frequency of 10 Hz \cite{Cho90,Hase03n,Hase12,Hase13}.
The excitation of GST films with a laser pulse of $E$ $\approx$ 1.53 eV generates nonequilibrium free carriers across the narrow band gap of $E_{g}$ $\approx$ 0.76 eV for the amo-GST and $E_{g}$ $\approx$ 0.39 eV for the fcc-GST \cite{Lee05}. 
The absorption coefficient ($\alpha$) of fcc-GST at 1.53 eV is $\alpha$ $\approx$ 5.7 $\times$ 10$^{-5}$ cm$^{-1}$, corresponding to an optical penetration depth $\xi$ $\approx$ 18 nm \cite{Lee05}.
On the other hand, the $\alpha$ of amo-GST at 1.53 eV is $\alpha$ $\approx$ 2.2 $\times$ 10$^{-5}$ cm$^{-1}$, corresponding to $\xi$ $\approx$ 45 nm \cite{Lee05}. 

\subsection{Acoustic phonon propagation}
The differences in the optical penetration depth enable us to investigate the effect of varying film thickness over the range $d$ = 20 nm to 80 nm. 
When $d > \xi$ the acoustic strain pulse propagates into the substrate while the reflected pulse at the sample surface is probed at the delay time of $2d/v$. The pulse shape is well separated from the thermal expansion component near $z$ = 0. The total strain is thus described by Thomsen's model \cite{Thomsen86},
\begin{eqnarray}
\eta (z, t) = \eta_{0} [e^{-z/\xi}(1- \frac{1}{2}e^{- vt/\xi}) - \frac{1}{2}e^{-|z - vt|/\xi} sgn(z - vt)], 
\end{eqnarray}
where $\eta_{0} = (1-R)\frac{I \beta}{\xi C_{l}}\frac{1 + \nu}{1 - \nu}$ is the amplitude of the strain, $R$ is sample reflectivity, $\nu$ is the Poisson's ratio, $C_{l}$ is the specific heat per unit volume, $\beta$ is the linear expansion coefficient \cite{Park08}, $I$ is the laser energy per area incident on the sample surface, and $v$ is the longitudinal sound velocity. The first term in Eq. (1) represents the strain due to thermal expansion, which dominates near $z$ = 0. The second part in Eq. (1) corresponds to the strain pulse which propagates away from the surface at the speed $v$ as shown in Fig. 1(a). 
\begin{figure}
\begin{center}
\includegraphics[width=7cm]{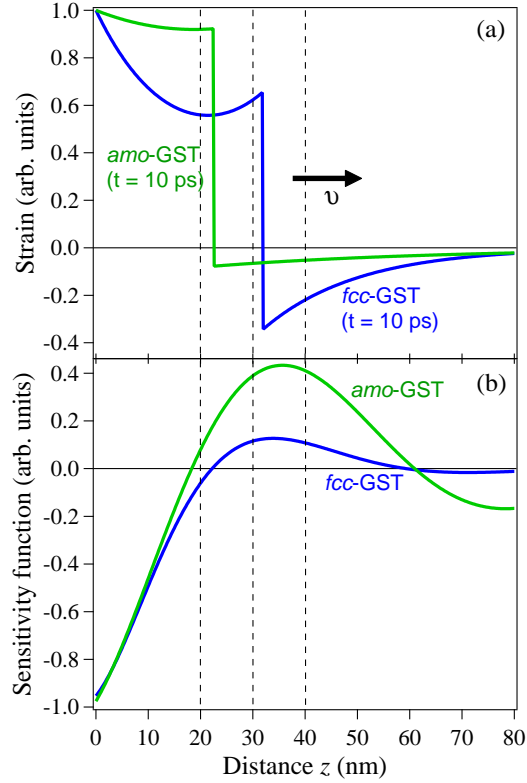}
\caption{(a) Simulated strains generated in amorphous and crystalline GST films by an optical pump pulse as a function of the distance from the surface ($z$) based on Thomsen's model, Eq. (1). The amplitude $\eta_{0}$ is set to unity for simplicity. The time delay $t$ is set to 10 ps after the pump pulse irradiation. The vertical dotted lines represent the positions of the film thickness used in the present study. (b) The sensitivity function $f(z)$ for the amorphous and crystalline GST films as a function of the distance from the surface based on the Thomsen's model, Eq. (2). The amplitude $f_{0}\frac{\partial \ln E_{g}}{\partial \eta}$ is set to unity for simplicity. 
The vertical dotted lines represent the position of the film thickness used in the present study. In the case of amo-GST, $f(z)$ has large amplitude till $z$ $\approx$ 45 nm. The values of $n$, $\xi$, and $\phi$ used are listed in Table I. 
}
\label{Fig1}
\end{center}
\end{figure}

The optical absorption of the laser pulse determines the factor of strain sensitivity $f(z)$ on the optical constants as a function of $z$ in response to a change in reflectivity and the sensitivity function $f(z)$ is given in terms of the penetration depth $\xi$ \cite{Thomsen86},
\begin{eqnarray}
f (z) = f_{0}\frac{\partial \ln E_{g}}{\partial \eta} \cos \Bigl[\frac{4\pi nz}{\lambda} - \phi \Bigr] \exp(-z/\xi),
\end{eqnarray}
where $f_{0}$ is the amplitude as a function of $n$ and $\kappa$, with $n$ and $\kappa$ being the real and imaginary part of the complex index of refraction, respectively \cite{Chabli02}, and $\phi = \tan^{-1}\Bigl[\frac{\kappa(n^{2} + \kappa^{2} + 1)}{n(n^{2} + \kappa^{2} - 1)}\Bigr]$ is the phase shift. For the case of the amorphous film ($\xi$ $\approx$ 45 nm), the pump beam penetrates into the sample beyond $z$ = 40 nm, and $f(z)$ has large amplitude till $z$ $\approx$ 45 nm, while for the case of fcc-GST ($\xi$ $\approx$ 18 nm) $f(z)$ immediately damped for $d$ $<$ 40 nm as shown in Fig. 1(b). There is thus a boundary where the condition changes from $d > \xi$ ($d$ = 80 nm amo-GST) to $d < \xi$ ($d$ = 20, 30, 40 nm amo-GST). If the condition $d < \xi$ is satisfied the pump light penetrates into the substrate (Si) and the entire GST film is excited. In this situation, the acoustic phonon oscillation approaches a state similar to a resonant quantum mode rather than a clearly separated acoustic strain pulse, whose wavelength is represented by $\Lambda$ = 2$d$ \cite{Thomsen87}. Then the frequency of this lowest eigenmode oscillation is given by $\omega = v/\Lambda = v/2d$, which in turn leads to a period of $T = 2d/v$. 

 \begin{table}
  \caption{\label{math-tab2}Parameters used in Eqs. (1), (2) and (5).  The values of $\phi$ were calculated by using $n$ and $\kappa$ based on the definition in the main text. 
    }
  \footnotesize\rm
  \begin{tabular*}{\textwidth}{@{}l*{15}{@{\extracolsep{0pt plus12pt}}l}}
\br
     Sample & $R$ (\%)$^{a}$ & $n$$^{a}$ & $\kappa$$^{a}$ & $\phi$ (radian) & $v$ (nm/ps)$^{b}$ & $\beta$ (K$^{-1}$)$^{c}$ & $\xi$ (nm)$^{d}$ \\
 \mr
  {\it amo}--Ge$_{2}$Sb$_{2}$Te$_{5}$ & 0.46  & 4.75 & 1.45 & 0.32 & 2.25 & 1.33$\times$10$^{-5}$ & 45  \\
  {\it fcc}--Ge$_{2}$Sb$_{2}$Te$_{5}$ & 0.59 & 5.45 & 3.31 & 0.57 & 3.19 & 1.74$\times$10$^{-5}$ & 18  \\
  \br
  \end{tabular*}\\
$^{a}$ from Ref. \cite{Chabli02}
$^{b}$ from Ref. \cite{Lyeo06}
$^{c}$ from Ref. \cite{Park08}
$^{d}$ from Ref. \cite{Lee05}
\end{table}

\section{Results and analysis}
\begin{figure}
\begin{center}
\includegraphics[width=8.0cm]{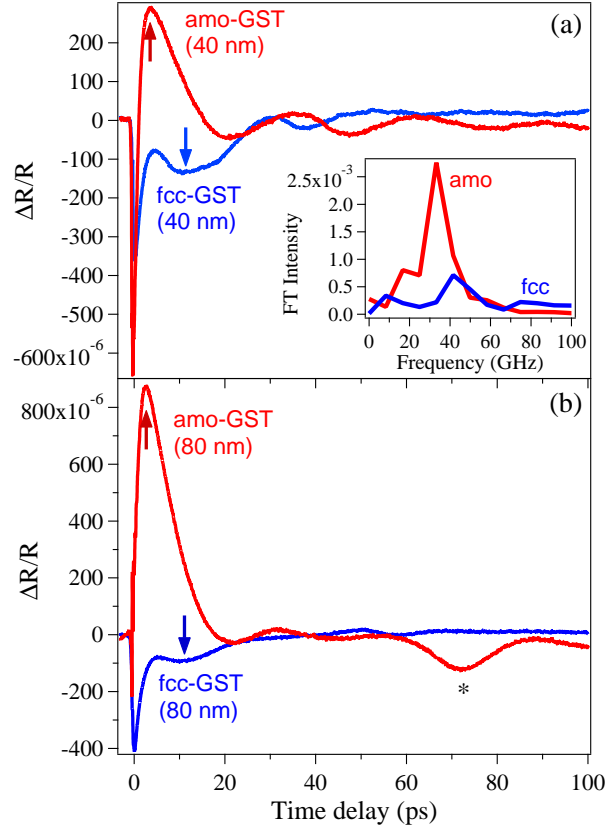}
\caption{The time-resolved TR signal observed in two different thickness GST films for the amorphous and crystalline phases pumped with a fluence of 5 mJ/cm$^{2}$.  (a) 40 nm thick and (b) 80 nm thick. The red and blue arrows represent the position of the acoustic response as a starting point. 
The inset  in (a) represents the FT spectra obtained from the time-domain data for the case of the 40 nm-thick films. The signal labeled by * in (b) represents the propagating acoustic strain pulse, which is separately detected. 
}
\label{Fig2}
\end{center}
\end{figure}
Figure 2 summarizes the time-resolved TR signal ($\Delta R/R$) observed in GST films at the two typical application thicknesses, 40 and 80 nm. 
After the transient electronic response (negative sharp dip) due to the excitation of nonequilibrium carriers at the time delay zero, relaxation of carriers due to carrier-phonon scattering with several picoseconds relaxation time is observed as the signal level recovers toward the original level. 
The time scale of the initial electronic response and the following carrier-phonon scattering is comparable to the previous studies \cite{Kim13,Shu13}.  
Just after the ultrafast electronic response, a slow acoustic signal starts to appear as an combination of an initial large exponential-like response, as indicated by the thick arrows, and following oscillatory response in the GHz frequency range (see the Fourier transformed spectra in the inset of Fig. 2a).  
The position of the initial large exponential-like response in the time delay (indicated by the thick arrows) does not change with the film thickness, i.e., $t$ = 2.5 $\pm$ 0.5 ps for amo-GST and 10.5 $\pm$ 0.5 ps for fcc-GST, while the time period of the oscillatory response depends on the film thickness, as discussed in detail later. 
Since the initial large exponential-like response indicated by thick arrows is film thickness independent, it should correspond to the strain near $z$ = 0, the first term in Eq. (1), while the oscillatory response can be explained either by a coherent acoustic phonon due to strain pulse propagating into and reflecting back from the Si substrate, the second term in Eq. (1), or an eigen-mode oscillation, depending on the condition of $d > \xi$ or $d < \xi$, respectively. 
The sign of the exponential-like strain is opposite for amorphous and crystalline cases, as observed in much thicker GST films \cite{Shu13}, indicating that the sign of the overall strain is opposite; expansion for the crystalline versus contraction for amorphous films upon photoexcitation. 

For the case of the $d$ = 40 nm amo-GST film, since the condition of $d < \xi$ is satisfied for the amorphous film, the acoustic phonon response appears as a single damped harmonic oscillation (see Fig. 3a) with the oscillation period of $T = 2d/v$ or the frequency of $\omega = v/2d$ \cite{Thomsen86},
\begin{eqnarray}
\Delta R (t) = R_{0}\exp\Bigl(-\frac{t}{\tau}\Bigr) \cos \Bigl[\frac{v}{2d}t - \delta \Bigr],
\end{eqnarray}
where $R_{0}$ is the amplitude, $\delta$ is the initial phase, and $\tau$ is the relaxation time of the acoustic phonon. 
\begin{figure}
\begin{center}
\includegraphics[width=7.5cm]{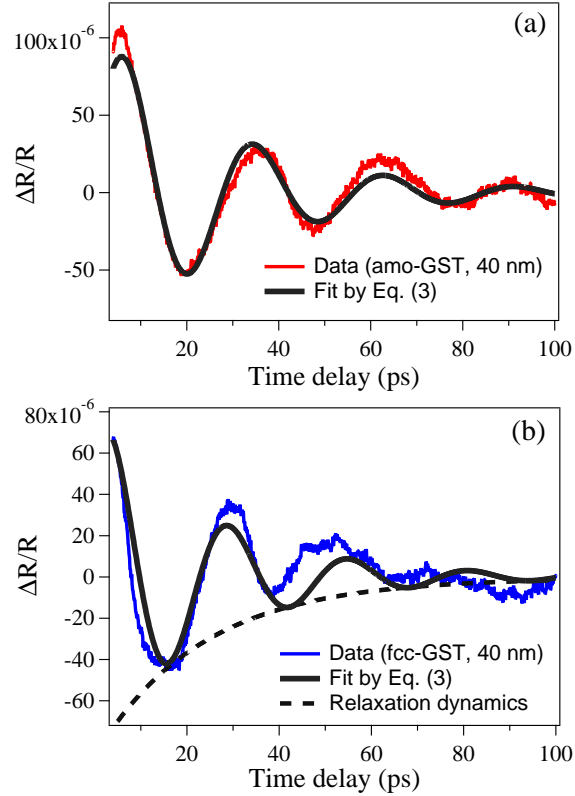}
\caption{(a) The acoustic phonon oscillation in amo-GST (40 nm) obtained by subtracting the electronic and the initial exponential-like responses from the time domain TR signal in Fig. 2(a). (b) The acoustic phonon oscillation in fcc-GST (40 nm) obtained from Fig. 2(a) in the same way. The black solid lines are the fit by Eq. (3) and the dashed line represents the envelope function for the acoustic strain pulse in order to estimate the time constant for the relaxation dynamics. 
}
\label{Fig3}
\end{center}
\end{figure}

\begin{figure}
\begin{center}
\includegraphics[width=7.0cm]{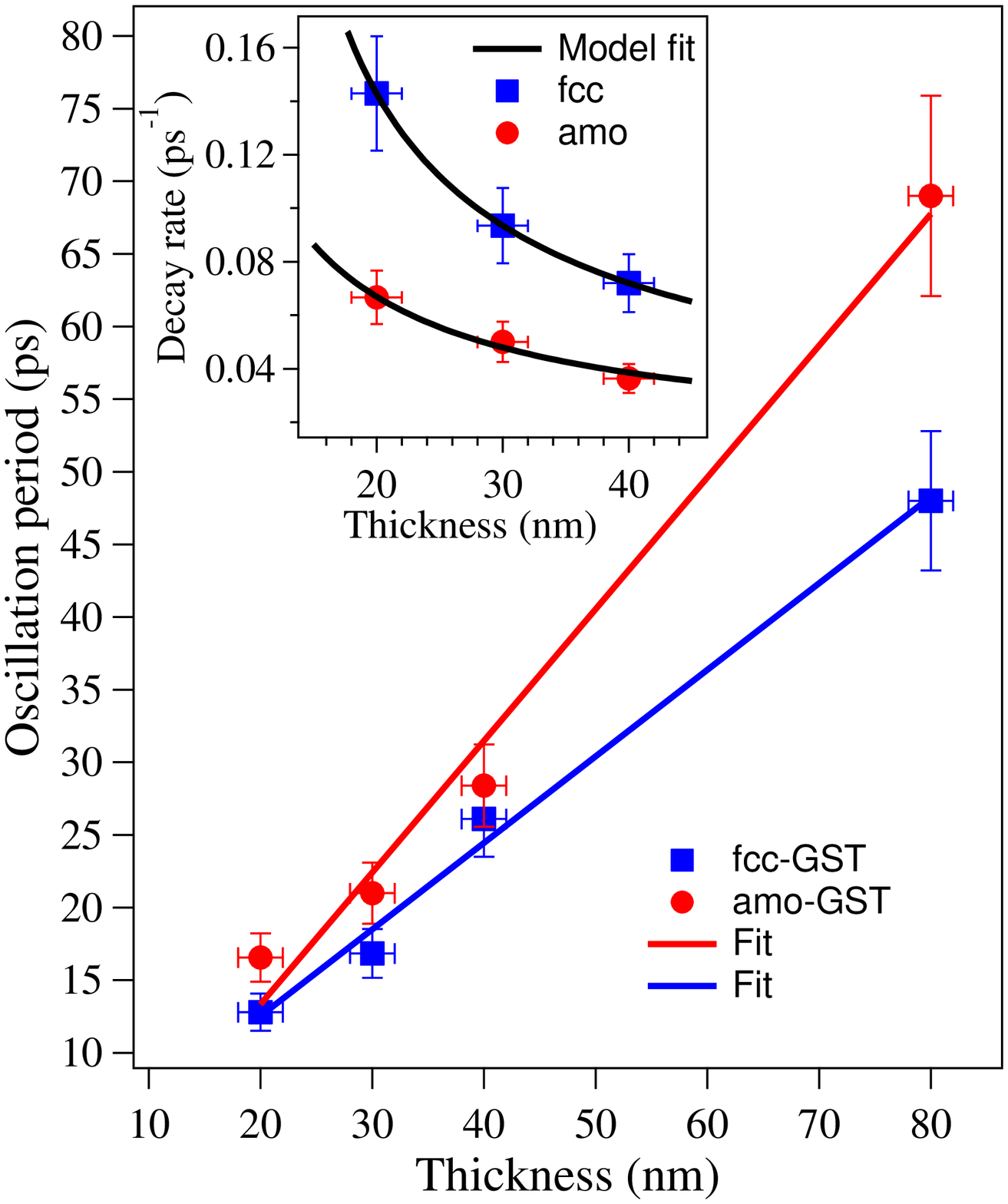}
\caption{The oscillation period of the acoustic phonon observed in the time-resolved TR signal of GST films at 5 mJ/cm$^{2}$ as a function of the film thickness. The oscillation period for the 80 nm-thick amo-GST and fcc-GST samples was obtained by reading the time difference between the point the acoustic response starts, indicated by thick arrows in Fig. 2 and the reflected strain pulse (* in Fig. 2). The solid lines are the fit with a linear function. The inset represents the decay rate ($\tau^{-1}$) of the acoustic phonon obtained by the fit of the time-domain data using Eq. (3). The solid curves in the inset are the model fit using Eq. (6) [see the main text in detail]. 
}
\label{Fig4}
\end{center}
\end{figure}
As shown in Fig. 3(a), we fit the acoustic phonon signal with Eq. (3), varying the oscillation period ($T$) and the relaxation time ($\tau$) for the case of a 40 nm-thick amorphous film ($\xi$ $\approx$ 45 nm, and therefore, $d < \xi$), and in addition the 20 nm and 30 nm-thick amorphous films (not shown). 
For the case of the crystalline films ($\xi$ $\approx$ 18 nm), on the other hand, since the condition of $d > \xi$ is always satisfied, the oscillatory response is observed rather as a  propagating strain pulse. Consequently, we cannot fit the acoustic phonon signal for the 20 - 40 nm-thick crystalline films satisfactory (see the case of the 40 nm film shown in Fig. 3b). Nevertheless, the envelope function, $R_{0}\exp(-t/\tau)$ in Eq. (3), is useful for estimating the time constant in fcc-GST films since it is acceptable to express the relaxation dynamics by the acoustic phonon envelope. 

When $d > \xi$ satisfied for a 80 nm-thick film, the acoustic phonon becomes a propagating acoustic strain pulse, as discussed below. Then the values of $T$ for the 80 nm-thick amo-GST and fcc-GST films are obtained by reading the time difference between the point at which the acoustic response starts, indicated by thick arrows in Fig. 2, and the minimum of the reflected pulse (see * in Fig. 2b). 
The values of the oscillation period ($T$) and the decay rate ($\tau^{-1}$) are obtained and plotted in Fig. 4 for various thicknesses. 
It is important to note that the time period of the acoustic oscillation does not depend on the pump fluence between 1.3 mJ/cm$^{2}$ and 5 mJ/cm$^{2}$ (not shown), values whose choice is made in order to prevent photo-bleaching effects and/or causing permanent phase change \cite{Rueda11,Takeda14,Fons14}, but it depends on the film thickness as shown in Fig. 4. 
The decay rate of the acoustic phonon oscillations also depends on the film thickness, as it decreases with increasing the film thickness from 20 to 40 nm (see Fig. 4 inset). 
Note also that the FT spectrum of amo-GST (40 nm) in the inset of Fig. 2(a) indicates the main peak exists at $\approx$ 33 GHz, approximately matching the acoustic frequency obtained by the inverse of the oscillation period of 28.4 ps (for amo-GST with 40 nm) as found in Fig. 4, i.e., 1/28.4 ps = 35 GHz. 
Thus we can estimate the value of the sound velocity $v$ by fitting the oscillation period with a linear function in Fig. 4 to be 3.36 nm/ps (3360 m/s) for fcc-GST and 2.21 nm/ps (2210 m/s) for amo-GST, matching well with those reported by Lyeo {\it et al} using thermo-reflectance \cite{Lyeo06} (see also Table I).

In the case of the thickest 80 nm GST film, the propagating acoustic strain pulse is separately detected (see * in Fig. 2b), which exhibits a shape well described by the Thomsen's model,\cite{Thomsen86} the second term in Eq. (1), as discussed below. 
Fig. 5 displays the acoustic strain pulse observed in the amorphous and crystalline films (both 80 nm-thick), whose shape is strongly bipolar \cite{Ren11}. A second echo pulse was not detected because of the limited time delay range up to 120 ps in the present study, although the second pulse echo is expected to appear at a later time delay of $>$120 ps. 
The data are fit to a model described by the convolution of the sensitivity function $f(z)$ [Eq. (2)] with the strain pulse, the second term in Eq. (1) \cite{Thomsen86}, 
\begin{eqnarray}
\Delta R (t) = \int_{0}^{d}f(z) \eta(z, t)\, dz.
\end{eqnarray}
The model fits the experimental data for amo-GST in Fig. 5(a) satisfactorily. 
\begin{figure}
\begin{center}
\includegraphics[width=7.0cm]{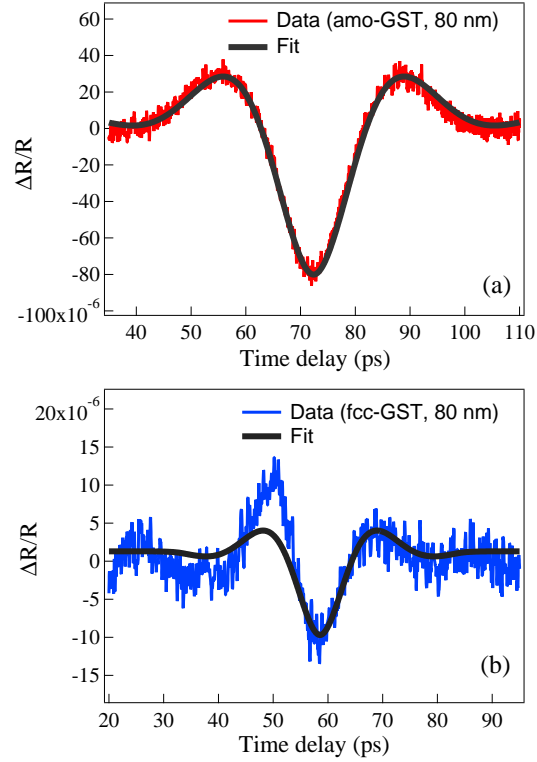}
\caption{The propagating acoustic strain pulse observed in the time-resolved TR signal of (a) amo-GST (80 nm) and (b) fcc-GST films (80 nm) at 5 mJ/cm$^{2}$, obtained by subtracting the background from the initial exponential-like response. The solid lines are a fit with the Thomsen's model (see the main text in details). 
}
\label{Fig5}
\end{center}
\end{figure}
In the case of the crystalline film, on the contrary, the model fits the experimental data for $t$ $>$ 55 ps, but there is discrepancy between them for $t$ $\leq$ 55 ps, which is due to the anti-symmetric shape. Such an anti-symmetric acoustic strain pulse has been observed in metal films and discussed in terms of the two temperature model (TTM) \cite{Wright94,Saito03}. Since the TTM can generally be applicable only for gap-less metals, we do not use the TTM for the current narrow-gap semiconductor GST film, since interband transitions dominate in the present case \cite{Lee05}. Hence the observed anti-symmetric shape in the crystalline GST is possibly due to remnants of the exponential-like response at $t$ = 10.5 ps, which cannot be completely subtracted away.

\section{Discussion}
\subsection{The magnitude of strain}
We discuss the difference in the acoustic phonon dynamics in the amorphous and crystalline GST films. 
The intensity of initial exponential-like strain at 2.5 ps for amo-GST is larger than that at 10.5 ps for fcc-GST as shown in Fig. 2 by a factor of $ \approx$ 2 in the case of 40 nm-thick films and by $ \approx$ 9 in the case of 80 nm-thick films. Since the magnitude of the strain is expressed by $\eta_{0} \propto (1-R)\frac{\beta}{\xi}$ for the same pump fluence according to Eq. (1), we can roughly estimate the ratio of the strain under the condition that the Poisson's ratio ($\nu$) \cite{Park08} and specific heat ($C_{l}$) \cite{Kuwahara07} do not change significantly for the two different phases. By choosing the values listed in Table I, one gets $\frac{\eta_{0}^{amo}}{\eta_{0}^{fcc}} \approx $ 0.3, a value not able to explain the larger amplitude of the light-induced strain observed in amo-GST. 

In order to account for the physical origin of the difference in the amplitude of the strain observed, it is useful to introduce a thermal stress related to lattice anharmonicity ($\sigma_{l}$) and a thermal pressure from hot electrons ($\sigma_{e}$) \cite{Thomsen86,Park05}. 
The faster time scale for the initial large exponential-like strain in the amorphous phase (2.5 ps) implies this response is dominated by the electronic stress ($\sigma_{e}$), as discussed by Shu {\it et al} \cite{Shu13}. On the other hand, the initial large exponential-like strain in the crystalline phase (10.5 ps) would be responsible for the thermal stress originating from lattice anharmonicity ($\sigma_{l}$). The ratio of $\sigma_{e}/\sigma_{l}$ is given by \cite{Thomsen86}, 
 \begin{eqnarray}
\frac{\sigma_{e}}{\sigma_{l}} = \frac{C_{l}}{3\beta(E - E_{g})}\frac{dE_{g}}{dP},
\end{eqnarray}
where $\frac{dE_{g}}{dP}$ denotes the change of the band-gap energy with pressure $P$. Putting $C_{l}$ = 1.35$\times$10$^{6}$J/m$^{3}$K \cite{Kuwahara07}, $\beta$ from Table I, $E - E_{g}$ = 0.76 eV for amo-GST and 1.14 eV for fcc-GST \cite{Lee05}, and $|\frac{dE_{g}}{dP}|$ = 0.014 eV/kbar for amo-GST \cite{Im10} and 0.004 eV/kbar for fcc-GST \cite{Xu11}, into Eq. (5) one obtain $\frac{\sigma_{e}}{\sigma_{l}}$ $\approx$ 6.23 for amo-GST and 0.91 for fcc-GST. The results indicate that  the contribution from the electronic stress ($\sigma_{e}$) dominates in amo-GST, while in fcc-GST the thermal stress ($\sigma_{l}$) plays a dominant role, being consistent with the interpretation given earlier. 

\subsection{Relaxation dynamics of coherent acoustic phonon}
The decay rate, the inverse of the relaxation time ($\tau^{-1}$), of the acoustic phonon is found to be larger in the case of the fcc-GST, as shown in the inset of Fig. 4. 
%Similar results have been obtained by other groups,\cite{Kim13,Shu13} although, here we discuss the effect in terms of film thickness. 
The damping of acoustic phonons is generally governed by scattering by disorders and/or scattering with underlying acoustic phonons in wavevector space \cite{Hase10}. 
The decrease in decay rate observed in Fig. 4 inset rather implies that scattering by disorders \cite{Hase10,Hase00} and attenuation at interfaces play roles in acoustic phonon relaxation in GST materials. 

In order to further discuss dynamics of the phonon scattering events in GST materials, the inverse of the acoustic phonon relaxation time, is introduced by using contributions from the phonon-defect scattering \cite{Pohl62,Wang09} and the attenuation at the GST/Si interface:
\begin{equation}
\tau^{-1} = A\omega^{4} + B\frac{v}{d} + \tau_{0}^{-1},
\end{equation}
where $A$ and $B$ characterize phonon-defect scattering and attenuation of acoustic phonon at the GST/Si interface, respectively. 
It is noted that the present study being operated at constant room temperature, all other terms depending on lattice temperature \cite{Pohl62,Wang09} is included as a constant, $\tau_{0}$. 
The relation between the oscillation period $T$ and the thickness $d$ shown in Fig. 4 gives $T_{fcc} = 0.59d$ and $T_{amo} = 0.91d$. 
We put the frequency $\omega_{fcc}(d)$ = $T_{fcc}^{-1}$ = $(0.59d)^{-1}$ into Eq. (6). In the same way, $\omega_{amo}(d)$ = $T_{amo}^{-1}$ = $(0.91d)^{-1}$ was input into Eq. (6). 
As shown in Fig. 4 inset, this model reproduces the thickness dependent relaxation time quite well, using fitting parameters of $A$ = 2.4 $\times$10$^{-46}s^{3}$, $B$ = 0.71 and $\tau_{0}^{-1}$ = 0.012 ps$^{-1}$ for fcc-GST and $A$ = 4.0 $\times$10$^{-47}s^{3}$, $B$ = 0.51 and $\tau_{0}^{-1}$ = 0.010 ps$^{-1}$ for amo-GST.  
From the parameters obtained, it is found from the magnitude of $A$ that the contribution from the phonon-defect scattering in fcc-GST is larger than amo-GST, and from the magnitude of $B$ that the attenuation of acoustic phonon at the fcc-GST/Si interface is stronger than that at the amo-GST/Si interface. 

It should be also mentioned that we observed a longer relaxation time of the coherent acoustic phonons in a film in the hexagonal crystalline phase (hcp-GST) \cite{Note2}, e.g., $\approx$ 28 ps for the 40 nm thick hcp-GST film, although the shape of $\Delta R/R$ signal and main acoustic parameter (the frequency and sound velocity) were quite similar to those in fcc-GST (not shown). This implies the disorder-induced phonon scattering is smaller in hcp-GST than fcc-GST possibly due to the ordered vacancies formed in hcp-GST, and a insulater-metal transition would occur upon the phase change from fcc-GST to hcp-GST films \cite{Zhang12}. 
Thus, the significantly defective nature of fcc-GST films would be relevant to disorder-induced electron (and even phonon) localization in crystalline phase change materials based on Ge, Sb, and Te, in which Anderson-like localization has been reported to occur \cite{Siegrist11,Zhang12}.

\subsection{Attenuation dynamics at interfaces}
The acoustic reflection coefficient $r$ for a perfectly flat interface is defined in terms of the impedance mismatch between the GST film and the Si substrate, $r = \frac{Z_{Si} - Z_{GST}}{Z_{Si} + Z_{GST}}$ \cite{Thomsen87}.  
Using the density $\rho$, we calculate $Z_{_{GST}} = \rho_{_{GST}} v_{_{GST}}$ = 6.4 g/cm$^{3}$ $\times$ 3360 m/s for fcc-GST, while $Z_{Si} = \rho_{_{Si}} v_{_{Si}}$ = 2.3 g/cm$^{3}$ $\times$ 9130 m/s for Si substrate \cite{Landolt}, yielding $r$ = 0.037 for the fcc-GST/Si interface. In the same way, we obtain $r$ = 0.227 for the amo-GST/Si interface. The results indicate that the acoustic attenuation at the GST/Si interface due to the impedance mismatch is stronger for the fcc-GST/Si interface, which is consistent with the experimental data in Fig. 5 inset and corresponding discussion. 
In order to control the degree of the acoustic attenuation at the GST/Si interface, at which $\sim$ 11 \% lattice mismatch is expected, epitaxial growth of GST thin films on GaSb substrate will be promising because the  lattice mismatch at the GST/GaSb interface is expected to be only $\sim$ 1 \% \cite{Rodenbach12}.

\section{Summary}
In summary, the picosecond dynamics of the coherent acoustic phonon response generated by femtosecond optical pulses in the two different phases of the prototypical phase change material, Ge$_{2}$Sb$_{2}$Te$_{5}$ film, has been investigated by using a pump-probe technique.
After excitation by near infrared pulse of 1.53 eV with $\approx$ 5 mJ/cm$^{2}$-fluence, the time-resolved reflectivity change exhibits transient electronic response, followed by the combination of an exponential-like strain and either an eigenmode or a traveling acoustic strain pulse in the several tens of GHz frequency range. Thomsen's model well describes the separated acoustic wave observed in the thickest GST films. Moreover, the ratio of electronic stress to the lattice thermal stress derived by application of Thomsen's model explains the larger exponential-like strain observed in amorphous GST films. The oscillation period of the acoustic phonon depends on the film thickness, from which the sound velocity of phase change materials was explicitly derived for the amorphous and crystalline phases of Ge$_{2}$Sb$_{2}$Te$_{5}$, matching well with the values reported by thermo-reflectance. In addition, the decay rate of the gigahertz acoustic phonons is found to be larger in the thicker GST films, which is well reproduced by taking into account both phonon-defect scattering and attenuation at the GST/Si interface. It is also demonstrated that both phonon-defect scattering and attenuation at the GST/Si interface are larger in fcc-GST than those in amo-GST, indicating the phonon localization effect in fcc-GST and the impedance mismatch at the fcc-GST/Si interface. 

Since disorder in the phase change materials can play a significant role in a possible metal-insulater transition, the different phonon scattering mechanisms observed in GST  films in the present study are useful to understand the physics of disorder-induced electron and phonon localization in prototypical GST materials. 

\ack{
This work was supported in part by X-ray Free Electron Laser Priority Strategy Program Projects (Nos. 12013011 and 12013023) from the Ministry of Education, Culture, Sports, Science and Technology (MEXT) of Japan. }

\end{document}